\documentclass[conference]{IEEEtran}
\usepackage{cite}
\usepackage{amsmath,amssymb,amsfonts}
\usepackage{algorithmic}
\usepackage{graphicx}
\usepackage{textcomp}
\usepackage{xcolor}
\usepackage{multirow}
\usepackage{pgfplots}
\usepackage{todonotes}
\usepackage{hyperref}
\usepackage{listings}

\usepackage[frozencache=true,cachedir=minted-cache]{minted}

\usepackage{booktabs}
\usepackage{caption}
\usepackage{subcaption}
\usepackage{cleveref}

\def\BibTeX{{\rm B\kern-.05em{\sc i\kern-.025em b}\kern-.08em
    T\kern-.1667em\lower.7ex\hbox{E}\kern-.125emX}}

\title{Evaluation of Programming Models and Performance for Stencil Computation on Current GPU Architectures}

\author{\IEEEauthorblockN{Baodi Shan}
\IEEEauthorblockA{TotalEnergies EP Research \& Technology US, LLC \\
Houston, Texas, USA\\
Stony Brook University \\
Stony Brook, New York, USA \\
baodi.shan@stonybrook.edu}
\and
\IEEEauthorblockN{Mauricio Araya-Polo}
\IEEEauthorblockA{TotalEnergies EP Research \& Technology US, LLC \\
Houston, Texas, USA}
}    

\begin{document}
\maketitle
\thispagestyle{plain}
\pagestyle{plain}


\begin{abstract}

Accelerated computing is widely used in high-performance computing. Therefore, it is crucial to experiment and discover how to better utilize GPUGPUs latest generations on relevant applications. In this paper, we present results and share insights about highly tuned stencil-based kernels for NVIDIA Ampere (A100) and Hopper (GH200) architectures. Performance results yield useful insights into the behavior of this type of algorithms for these new accelerators. This knowledge can be leveraged by many scientific applications which involve stencils computations.
Further, evaluation of three different programming models: CUDA, OpenACC, and OpenMP target offloading is conducted on aforementioned accelerators. We extensively study the performance and portability of various kernels under each programming model and provide corresponding optimization recommendations. Furthermore, we compare the performance of different programming models on the mentioned architectures. Up to $58\%$ performance improvement was achieved against the previous GPGPU's architecture generation for an highly optimized kernel of the same class, and up to $42\%$ for all classes. In terms of programming models, and keeping portability in mind, optimized OpenACC implementation outperforms OpenMP implementation by $33\%$. If portability is not a factor, our best tuned CUDA implementation outperforms the optimized OpenACC one by $2.1\times$.

\end{abstract}

\section{Introduction}

Currently, HPC-based on General Purpose Graphic Processing Units (GPGPUs) is mostly replacing Central Processing Units (CPUs)-based computing, thus becoming the primary source of computational power for supercomputing systems. In the latest TOP500 supercomputer rankings released in Nov. 2023, fifteen out of the twenty fastest systems use GPUs as accelerators for supercomputing. 
Therefore is relevant to continue updating and optimizing workloads that rely on this kind of accelerators. 

The evolution of GPGPUs architectures is driven by different markets/applications needs, it is unknown which and to what extent the new features might be useful for specific numerical workflows, this is another reason to evaluate how well-established workflows map to new hardware. In this work's case, the target application is subsurface characterization (\cite{tylor2023practical}), through wave equation solving, which at the core sports high-order stencil computations. Research on stencil computing continuously produces technical advances (\cite{sun2023adapting, denzler2023casper, sun2022stencilmart, jacquelin2022scalable}) given its importance for many scientific and industrial applications(\cite{anshu14}), from weather prediction (\cite{fuhrer_towards_2014}) to earthquake modeling (\cite{moczo2014}).   

In this work, through performance evaluation under different programming models and data sizes, profiling analysis, and comparison of hardware parameters between GPGPUs, we provide suggestions for developers. Thus, this work has the following main contributions:
\begin{itemize}
    \item By evaluating our stencil computation program on the NVIDIA GH200, we explored the potential impact of the new feature of CUDA programming model based on the latest generation of NVIDIA GPGPU, thread block cluster, and analyzed the causes of performance changes.
    \item Based on the NVIDIA Hopper architecture, we proposed optimizations for the stencil computation program and provided optimization suggestions for subsequent developers.
    \item On the comparison between OpenACC's \texttt{async} and OpenMP's \texttt{nowait}, we proposed a new optimization strategy for asynchronously executing parallel regions, enabling the code to fully utilize the ability of the GPGPUs multiple streams to execute concurrently. At the same time, we also proposed optimization strategies at the compilation level for the stencil computation program on the OpenACC programming model. Compared with the original code, the performance of the new code has been improved by up to 30\%.
    \item We compared the performance and portability of three different GPGPU-based programming models, and evaluated and analysed the performance and changes of the three models on different generations of NVIDIA GPGPUs. Based on our evaluation results and analysis, we have provided corresponding suggestions to developers of scientific programs.
    \item We compared the power consumption of NVIDIA Ampere architecture and Hopper under three different programming models. This reflected the relationship between energy consumption and performance.
\end{itemize}

The paper is organized as follows:~\Cref{sec:impl} introduces the concept of stencil computation.~\Cref{sec:setup} describes the system overview and programming models used in our evaluations. \Cref{sec:cuda} showcases various implementations and optimization for the stencil computation program under the CUDA programming model, compares their performance on A100 and GH200, and provides optimization recommendations based on these results.
\Cref{sec:acc-omp} elaborates on the implementation of the the stencil computation program using OpenACC and OpenMP, presents our newest optimization scheme and its performance on the A100 and GH200.
\Cref{sec:gene} compared the portability of three different programming models, evaluated them on three generations of NVIDIA GPU's architectures and provides recommendations for GPU program developers on selecting programming models, considering both program portability and performance differences on NVIDIA Hopper architecture.

    

\section{Stencil Computation and Related Work}
\label{sec:impl}

The stencil pattern under analysis in this work helps to compute the differential operators required by Finite Difference (FD) scheme to solve an acoustic isotropic approximation of the wave equation. The base implementation was introduced in \cite{meng2020minimod} and optimized versions tailored for GPGPUs in presented in \cite{pmbs20}. The spatial part of the wave equation is discretized using a 25-point stencil in 3D ($8^{th}$ order in space), with four points in each direction as well as the centre point:
\begin{align*}
\nabla^2 \mathbf{u}(x,y,z) \approx \sum_{m=0}^4 &c_{xm}\left[\mathbf{u}(i+m,j,k) + \mathbf{u}(i-m,j,k)\right] &+ \\
									 &c_{ym}\left[\mathbf{u}(i,j+m,k) + \mathbf{u}(i,j-m,k)\right] &+ \\
									 &c_{zm}\left[\mathbf{u}(i,j,k+m) + \mathbf{u}(i,j,k-m)\right] \tag{1}
\end{align*}
where $c_{xm},c_{ym},c_{zm}$ are discretization parameters and $\mathbf{u}$ the wavefield.

Basically, the computing pattern in Equation 1 is computed per every grid point of the computational domain and used to update the wavefield per time iteration. For instance, for a 3D computing domain of size $1000^3$, the pattern in Equation 1 is computed 1E09 times per time step, if the simulation iterate $1000$ time steps, then pattern is computed 1E12 times in total. 
Thus, Equation 1 represents the more computationally challenging step of solving the wave equation (by FD) since the memory access pattern overwhelms traditional memory hierarchies, due to its low re-use, and the sparse in-memory location of the required elements to compute the central point of the stencil. Further, the computational domain is surrounded by CPML-like (\cite{koma2007}) boundary condition, which is implemented as part of the inner domain loops.  

Multiple works introduce implementations and optimization strategies for stencil computations, relevant to our work references are described as follows. The strategy of overlapped tiling employs time skewing to amplify the arithmetic intensity of parallel stencil computations. This is done by exchanging redundant computation along the boundaries of overlapped tiles for a decrease in the required memory bandwidth~\cite{10.1145/1250734.1250761},~\cite{10.1145/2304576.2304619}. This approach is particularly effective on GPUs due to the high cost of loading data from a GPU’s global memory compared to data-parallel computation. Moreover, redundant computation can be overlapped with data accesses to help conceal memory latency. While overlapped tiling has been demonstrated to enhance the performance of low-order stencils on GPUs, for high-order stencils, the redundant computation escalates rapidly when skewed across multiple time steps by the width of a high-order stencil.
The time skewing approach, as discussed in~\cite{845979},~\cite{Skewing}, enhances the performance of stencil computations by augmenting data reuse and cache locality. This is achieved by skewing one or more data dimensions by the time dimension, enabling the computation of several time steps for a tile while the values are retained in cache. This method has found extensive application in CPUs, as evidenced by~\cite{jin2001increasing} and~\cite{10.1145/301618.301668}.
An alternative strategy for accelerating computation with time skewing is split tiling, as described in~\cite{10.1145/2458523.2458526}. Instead of using overlapped tiles, which can induce significant amounts of redundant computation, split tiling computes points in two distinct phases. The initial phase computes tiles in parallel as hypertrapezoids that taper along the time dimension. Once all tiles from the first phase have been computed, a second phase fills in the missing points in the time dimension.

Nguyen et al.~\cite{nguyen20103} propose a 3.5D blocking algorithm that blends 2.5D spatial blocking with 1D temporal blocking. 2.5D spatial blocking involves blocking in a 2D plane and streaming along a third dimension to increase data reuse, storing active 2D planes in GPU shared memory. In the 3.5D variant, they use time skewing to advance the computation for multiple time steps before writing data back to the global memory. Although the 3.5D algorithm performs exceptionally well on CPUs, the 1D temporal blocking introduces two potential implementation challenges for high-order stencils with boundary conditions on GPUs: barrier synchronizations and limited parallelism.

\section{Evaluation Setup}
\label{sec:setup}
\subsection{System Overview}

The primary testing platform is a NVIDIA Grace Hopper Superchip with NVIDIA GH200. In the comparative evaluations, we also utilized computing nodes equipped with A100 and T4. Refer to~\Cref{table:machine-spec} for the hardware and software specifications of the systems.

\begin{table}[t]
\small
    \centering
    \caption{System Specifications}
    \label{table:machine-spec}
    \resizebox{\linewidth}{!}{%
    \begin{tabular}{|c | c | c | c |}
    \hline
        & GH200 & A100  & T4\\
    \hline
    CPU & NVIDIA Grace & AMD EPYC 7F52 & Intel(R) Xeon(R) 5217 \\
    \hline
    GPU & NVIDIA GH200 & NVIDIA A100 & NVIDIA T4 \\
    \hline
    Cores & 16896   & 6912 & 2560\\
    \hline
    GRAM & 96 GB & 40 GB & 16 GB\\
    \hline
    Memory Bandwidth & 4TB/s & 1555 GB/s & 320 GB/s \\
    \hline
    CUDA Compiler & 12.2 & \multicolumn{2}{|c|}{12.0} \\
    \hline
    OpenACC Compiler &  \multicolumn{3}{|c|}{NVHPC (nvc)} \\
    \hline
    OpenMP Compiler &  \multicolumn{3}{|c|}{NVHPC (nvc)} \\
    \hline
    GPU Driver & 535.129.03 & \multicolumn{2}{|c|}{525.105.17} \\
    \hline
    Compiler Arch & sm\_90 & sm\_80 & sm\_75 \\
    \hline
    \end{tabular}}
    
\end{table}

\subsection{New in NVIDIA Hopper Architecture}

\begin{figure}
     \centering
     \begin{subfigure}[b]{.7\linewidth}
         \centering
         \includegraphics[width=\linewidth]{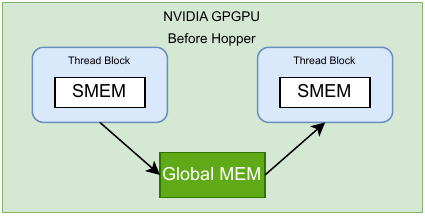}
         \caption{NVIDIA GPGPU without thread block clusters}
         \label{NVIDIA GPGPU before H100}
     \end{subfigure}
     \hfill
     \hfill
     \begin{subfigure}[b]{.7\linewidth}
         \centering
         \includegraphics[width=\linewidth]{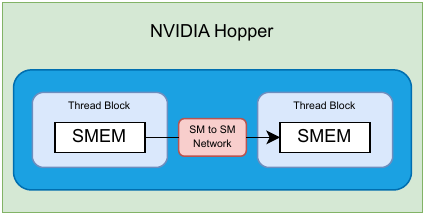}
         \caption{NVIDIA Hopper with thread block clusters}
         \label{NVIDIA H100}
     \end{subfigure}
        \caption{Thread-block-to-thread-block data exchange}
        \label{fig:tb_data_exchange}
\end{figure}

The NVIDIA Hopper Architecture introduces a new memory indexing layer, termed ``thread block cluster'', which needs to be analyzed properly. The motivation of this feature is to facilitates data locality control at a granular level exceeding a solitary thread block on an individual Streaming Multiprocessor (SM). It augments the CUDA programming model, incorporating an extra hierarchical level, which includes threads, thread blocks, thread block clusters, and grids. Thread block clusters facilitate simultaneous execution of numerous thread blocks over several SMs, fostering synchronization, and cooperative data retrieval and exchange. Figure \ref{fig:tb_data_exchange} illustrates various data exchange schemes occurring in NVIDIA GPGPUs, both with and without thread block clusters.

\subsection{Programming Models}

\textbf{CUDA} is a parallel computing platform by NVIDIA\cite{CUDA}. It allows developers to use a CUDA-enabled GPU for general purpose processing. In CUDA programming, developers write functions, known as ``kernel'', in a language similar to C/C++, which are then executed on the GPU for faster computations, especially beneficial for tasks like matrix operations, image processing, and machine learning.

\textbf{OpenACC} is a programming standard designed to simplify parallel computing, particularly for programming with accelerators like GPUs. It offers a directive-based programming model that lets developers indicate which parts of the code should run in parallel by inserting directives, without needing to specify how to parallelize in detail. The primary goal of OpenACC (\cite{OpenACC}) is to provide a simple way to optimize and parallelize applications with minimal effort from the developer.

\textbf{OpenMP} is an application programming interface (API) that supports multi-platform shared memory multiprocessing programming in C, C++, and Fortran. OpenMP (\cite{OpenMP,9741290}) provides a simple and flexible interface for developing parallel applications for platforms ranging from desktop computers to supercomputers. OpenMP's \texttt{target} directive is part of its support for offloading computation to accelerators, such as GPUs ~\cite{iwomp22,omptarget,10.1007/978-3-030-85262-7_11, 10.1007/978-3-031-40744-4_12}. Currently, in the implementation of OpenMP in LLVM, the OpenMP target offloading support NVIDIA GPU, AMD GPU, Intel Phi and remote devices\cite{pmam23}.


\section{Exploration and Optimization of CUDA Implementations}
\label{sec:cuda}
The first GPU programming model we explore is CUDA. Our program encompasses a diverse array of kernel implementations, each of which has been extensively described in Sai et al.\cite{pmbs20}. In the present document, we confine our discussion to their capacity to accommodate to nuanced idiosyncrasies inherent to distinct generations of GPGPUs.
We selected the best kernels in-class from it. The selection criteria and detailed description of these CUDA kernels will be discussed in ~\autoref{sec:cuda-kernel}

This section elaborates on the following aspects: Firstly, we explore several implementations of kernels under the CUDA model. Then, we present the performance and profiling results of each kernel implementation on NVIDIA Ampere and NVIDIA Hopper Architectures. Next, we discuss the following issues separately and provide corresponding suggestions: 1. The impact of the newly introduced Thread Block Cluster on program performance on NVIDIA Hopper and its causes. 2. The differences in the performance of the best kernels on NVIDIA Ampere and NVIDIA Hopper Architectures and their causes.

\subsection{Evaluated CUDA Kernels}
\label{sec:cuda-kernel}
As mentioned in the~\autoref{sec:impl}, we will focus on analyzing the following four implementations: \textit{gmem}, \textit{smem}, \textit{st\_reg\_fixed}, and \textit{st\_semi}.

\begin{figure}
\centering
\includegraphics[width=.6\linewidth]{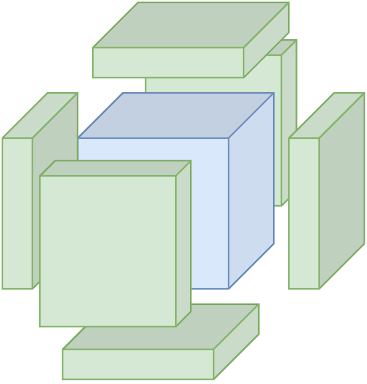}
  \caption{Data domain decomposition.}
  \label{fig:domain}
\end{figure}

As delineated in~\Cref{sec:impl}, our data domain encompasses two distinct regions: the inner region (computational region) and the Perfectly Matched Layer (PML) boundaries. The inner region constitutes a cubic grid situated at the core of the data domain, while the PML region signifies the volume interposed between the inner region and the boundaries of the data domain.~\Cref{fig:domain} shows the data domain decomposition strategy in our stencil computation program. In our discussion of implementation strategies and performance characteristics on the CUDA model, our primary focus is on the performance of the inner region.

In our various implementations, we primarily include the following strategies. The first is ``3D Blocking Using Global Memory Only'' which is denoted as \textit{gmem}; the second strategy is ``3D Blocking Using Shared Memory,'' which is denoted as \textit{smem}; the third strategy is `` 2.5D Streaming Fixed Registers,'' which we denote as \textit{st\_reg\_fixed}; the fourth strategy is `` 2.5D Streaming Semi-stencil,'' which was initially introduced for CPUs (\cite{10.1145/2591006}), which purpose is to increase memory reuse, it is denoted as \textit{st\_semi}.

\begin{figure}
\centering
\includegraphics[width=\linewidth]{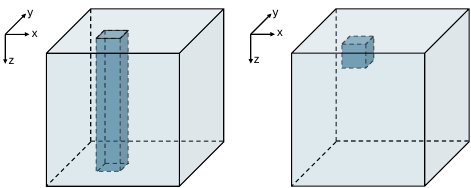}
  \caption{Blocking strategies: (left) 3D blocking, (right) 2.5D blocking.}
  \label{fig:3d2.5d}
\end{figure}

In order to better illustrate the different optimization strategies, we first introduce the two blocking strategies adopted in our stencil computation program. Our stencil computation program essentially deals with a 3D problem, so the basic blocking strategy is 3D blocking. In 3D Blocking, each data region is segmented into 3D blocks aligned with the axis. In order to determine the optimal block dimensions, we use fixed values in each run to simplify trials with varying values. For stencil computations on GPUs, each block is associated with a kernel launch with a 3D thread block, the thread dimensions of which match the block dimensions. All points within the block and their halos are explicitly copied into the GPU's on-chip memory prior to any kernel initiation. While in 2.5D blocking strategy, we divide the data domain along the inner two dimensions, X and Y, and conduct a streaming computation along the outermost Z dimension. We initiate kernels with 2D thread blocks, the dimensions of which correspond to the 2D planes. Difference of these two blocking strategies could be seen in ~\autoref{fig:3d2.5d}.

In our kernel descriptions, the symbol $R$ represents the width of the halo, which is half the spatial order of the stencil. In the acoustic isotropic simulations conducted in our experiments, the value of $R$ is set to 4. Additionally, let $N_x$, $N_y$, and $N_z$ denote the extents of the input data region along the $X$, $Y$, and $Z$ axes, respectively. In the context of 3D blocks, the notation $(x, y, z)$ is used to indicate the 3D coordinates, representing both a point location within a 3D block and the thread within a kernel thread block. Similarly, for 2D planes, the notation $(x, y)$ is employed to specify a point in the 2D plane and identify the corresponding thread.

In the mentioned strategies, \textit{gmem} means 3D blocking using global memory only. During the execution of the 25-point stencil kernel, each thread retrieves its own point and 4 neighboring points along each direction of every axis. To optimize performance, we ensure a favorable memory access pattern when fetching stencil points directly from global memory. This is achieved by storing the 3D grid data as a flat 1D array. Specifically, we prioritize global memory coalescing for the innermost dimension, $X$, to enhance data retrieval efficiency.

The \textit{smem} is utilized to denote the methodology of 3D blocking via shared memory specifically for the array $u$, the definition of which is explicated in Equation 1. This strategy is a modification of the previously discussed 3D blocking employing global memory. It retains the same 3D blocking strategy for each data region. However, in contrast to performing computations directly on data procured from the global memory, this approach retrieves the $u$ array from the global memory, stores it in the shared memory, and conducts the stencil computation on data procured from the shared memory. The total quantity of points procured in this scenario is $D_x \times D_y \times D_z$ for a block and $(D_x \times D_y + D_x \times D_z + D_y \times D_z) \times R \times 2$ for halos surrounding the block. In the context of high order stencils, it is imperative to consider the halo size to ascertain that both the block and the halo are accommodated within the shared memory.

Pertaining to the ``Fixed Registers'' strategy, it represents a variant from the ``Shifting Registers'' approach, a 2.5D streaming approach. In the ``Shifting Registers'' strategy, we maintain the points of the current $XY$-subplane in shared memory. However, as we stream along the z-axis, we utilize registers for the points along the z-axis. Unlike shared memory, where data procured from one thread is accessible by other threads within the same block, registers are solely accessible by the current thread. Given that we are streaming along the z-axis, the data from the z-axis loaded for one thread is not required by other threads. Consequently, we allocate a shared memory space to accommodate $(Dx + 2R) \times (Dy + 2R)$ points for the currently active plane. The shared memory footprint in comparison to the previous method is $1 : (2R + 1)$. For high-order stencils, $R$ is large, thereby leading to a significant reduction in shared memory usage.
However, in the ``Fixed Registers'' strategy, we ascertain that the values in the registers remain ``fixed'' rather than ``shifted'' and update the necessary data via read and write operations between shared memory and global memory.

As for ``Semi-stencil'' \cite{10.1145/1250734.1250761}, it separates the computation into two phases: forward and backward. The algorithm reads $R + 1$ points in one dimension and performs calculations as if they were the left side of the stencil in the forward phase, storing partial results. In the backward phase, the points are treated as the right side of the stencil, and the final result is written back. This approach changes the load-to-store ratio, reducing the number of loads. This implementation further integrates with a 2.5D streaming approach.

\subsection{Performance Results and Profiling on CUDA Kernels}

\begin{table*}[!h]
\centering
\caption{Execution time for selective CUDA kernels on NVIDIA Ampere, NVIDIA Hopper without and with thread block cluster. The numbers in angle brackets indicate the dimension of the thread block cluster. Figures in bold font represents the best performance in each column. The number in parentheses represents the performance improvement ratio of the optimal kernel in the column compared to the optimal kernel of A100.}
\label{cuda-time}
\resizebox{.78\linewidth}{!}{%
\begin{tabular}{@{}cccc@{}}
\toprule
\textbf{Kernel}                          & \textbf{A100 [s]}                    & \textbf{GH200,without thread block cluster [s]} & \textbf{GH200, with thread block cluster [s]} \\ \midrule
\multirow{2}{*}{gmem}           & \multirow{2}{*}{27.058} & \multirow{2}{*}{11.710}          & \textless1,3,1\textgreater12.041                   \\
                                &                         &                                   & \textless1,1,3\textgreater11.315                   \\ \midrule
\multirow{2}{*}{smem}           & \multirow{2}{*}{23.740} & \multirow{2}{*}{11.635}          & \textless1,3,1\textgreater12.300                   \\
                                &                         &                                   & \textless1,1,3\textgreater11.751                   \\ \midrule
\multirow{4}{*}{st\_reg\_fixed} & \multirow{4}{*}{19.908} & \multirow{4}{*}{\textcolor{red}{\textbf{9.445(41.5\%)}}}          & \textcolor{red}{\textbf{\textless1,2\textgreater9.434(41.6\%)}}                     \\
                                &                         &                                   & \textless1,4\textgreater9.733                      \\
                                &                         &                                   & \textless2,1\textgreater9.455                      \\
                                &                         &                                   & \textless4,1\textgreater9.776                      \\ \midrule
\multirow{4}{*}{st\_semi}       & \multirow{4}{*}{\textcolor{red}{\textbf{16.154}}} & \multirow{4}{*}{9.654}          & \textless1,2\textgreater10.435                     \\
                                &                         &                                   & \textless1,4\textgreater10.757                     \\
                                &                         &                                   & \textless2,1\textgreater10.799                     \\
                                &                         &                                   & \textless4,1\textgreater11.585                     \\ \bottomrule
\end{tabular}
}
\end{table*}

The grid size used for all the experiments in this section is $1024^3$ and the number of time iterations is $1000$.
\autoref{cuda-time} shows execution time for different CUDA kernels
, due to the requirement that the size of thread block cluster dimension must be a multiple of the GPU block in the corresponding dimension, we have chosen different dimension values for different CUDA kernels.

\begin{table*}[h]
\centering
\caption{Profiling results of memory information of CUDA kernels on A100 and GH200. The numbers in the brackets represent the percentage change in memory throughput and device memory between adjacent data.}
\label{tab:cuda-prof-mem}
\resizebox{\textwidth}{!}{%

\begin{tabular}{@{}ccccccc@{}}
\toprule
\multicolumn{1}{c}{\textbf{Kernel}}                             & \textbf{Device}    & \textbf{Memory Throughput Rate} & \textbf{Memory Throughput(GB/s)} & \textbf{L1 Hit Rate} & \textbf{L2 Hit Rate} & \textbf{Device Memory Read/Write(GB)} \\ \midrule
\multirow{3}{*}{gmem}           & A100                                & 52.45\%                                          & 815.59                                            & 74.95\%                               & 47.01\%                               & 22.15                                                  \\
                                & GH200                               & 59.57\%                                          & 2400.00                                           & 74.91\%                               & 33.05\%                               & 22.14                                                  \\
                                & GH200-\textit{opt} & 44.67\%                                          & 1800.00                                           & 75.11\%                               & 44.67\%                               & \textbf{17.23(-22.18\%)}                                        \\ \midrule
\multirow{3}{*}{smem}           & A100                                & 60.88\%                                          & 946.96                                            & 11.64\%                               & 47.28\%                               & 22.44                                                  \\
                                & GH200                               & 55.57\%                                          & 2240                                              & 11.61\%                               & 33.45\%                               & 22.42                                                  \\
                                & GH200-\textit{opt} & 40.65\%                                          & 1640                                              & 12.20\%                               & 46.30\%                               & \textbf{17.42(-22.31\%)}                                        \\ \midrule
\multirow{3}{*}{st\_reg\_fixed} & A100                                & 39.87\%                                          & 621.87                                            & 32.09\%                               & 56.07\%                               & 34.40                                                  \\
                                & GH200                               & 27.80\%                                          & 1110                                              & 27.98\%                               & 59.36\%                               & 23.55                                                  \\
                                & GH200-\textit{opt} & 27.28\%                                          & 1100                                              & 27.28\%                               & 60.59\%                               & \textbf{23.06(-2.08\%)}                                                  \\ \midrule
\multirow{3}{*}{st\_semi}       & A100                                & 33.53\%                                          & 519.70                                            & 33.46\%                               & 52.36\%                               & 26.12                                                  \\
                                & GH200                               & 23.63\%                                          & 950.54                                            & 34.46\%                               & 46.71\%                               & 22.39                                                  \\
                                & GH200-\textit{opt} & 23.26\%                                          & 935.73                                            & 34.47\%                               & 48.16\%                               & \textbf{21.91(-2.14\%})                                                \\ 
\midrule
\end{tabular}
}
\end{table*}

\begin{table*}[]
\centering
\caption{Profiling results of compute information of CUDA kernels on A100 and GH200.}
\label{tab:cuda-prof-comp}
\resizebox{\textwidth}{!}{

\begin{tabular}{@{}cccrrcccc@{}}
\toprule
\multirow{2}{*}{\textbf{Kernel}} & \multirow{2}{*}{\textbf{Device}}    & \multirow{2}{*}{\textbf{Compute Throughput}} & \multirow{2}{*}{\textbf{SM Frequency}} & \multirow{2}{*}{\textbf{Elapsed Cycles}} & \multicolumn{2}{c}{\textbf{Theoretical Occupancy}} & \multicolumn{2}{c}{\textbf{Achieved}} \\ \cmidrule(l){6-9} 
                                 &                                     &                                              &                                        &                                          & Occupancy           & Active Warps Per SM          & Occupancy    & Active Warps Per SM    \\ \midrule
\multirow{3}{*}{gmem}            & A100                                & 57.26\%                                      & 764,838,253                            & 20,763,880                               & 100                 & 64                           & 88           & 56                     \\
                                 & GH200                               & 77.86\%                                      & 1,528,630,472                          & 14,133,370                               & 100                 & 64                           & 82           & 52                     \\
                                 & GH200-\textit{opt} & 75.05\%                                      & 1,528,421,734                          & 14,666,940                               & 100                 & 64                           & 76           & 49                     \\ \midrule
\multirow{3}{*}{smem}            & A100                                & 67.21\%                                      & 765,095,453                            & 18,110,315                               & 100                 & 64                           & 92           & 59                     \\
                                 & GH200                               & 71.14\%                                      & 1,529,208,988                          & 15,339,753                               & 100                 & 64                           & 87           & 55                     \\
                                 & GH200-\textit{opt} & 67.00\%                                      & 1,527,931,405                          & 16,294,783                               & 100                 & 64                           & 81           & 52                     \\ \midrule
\multirow{3}{*}{st\_reg\_fixed}  & A100                                & 32.29\%                                      & 767,236,890                            & 42,435,694                               & 50                  & 32                           & 47           & 30                     \\
                                 & GH200                               & 36.22\%                                      & 1,529,787,652                          & 32,207,850                               & 50                  & 32                           & 47           & 30                     \\
                                 & GH200-\textit{opt} & 36.29\%                                      & 1,529,783,280                          & 32,144,449                               & 50                  & 32                           & 46           & 30                     \\ \midrule
\multirow{3}{*}{st\_semi}        & A100                                & 34.85\%                                      & 762,292,692                            & 38,308,107                               & 50                  & 32                           & 48           & 31                     \\
                                 & GH200                               & 31.58\%                                      & 1,529,861,717                          & 36,041,202                               & 50                  & 32                           & 49           & 31                     \\
                                 & GH200-\textit{opt} & 31.77\%                                      & 1,529,839,704                          & 35,830,759                               & 50                  & 32                           & 48           & 31                     \\ \bottomrule
\end{tabular}
}
\end{table*}

\autoref{tab:cuda-prof-mem} and \autoref{tab:cuda-prof-comp} show profiling results, the tables focus on memory and computation information respectively.
We selected the fastest version with the thread block cluster feature under each CUDA implementation and marked it as ``\textit{opt}''.

We will focus on the relationship between changes in GPU memory read and write volume and throughput in ~\autoref{sec:cluster}, we specifically annotated the changes in read and write volume and throughput brought about by the thread block cluster feature and represented these changes in parentheses. 

\begin{figure}
\centering
\includegraphics[width=\linewidth]{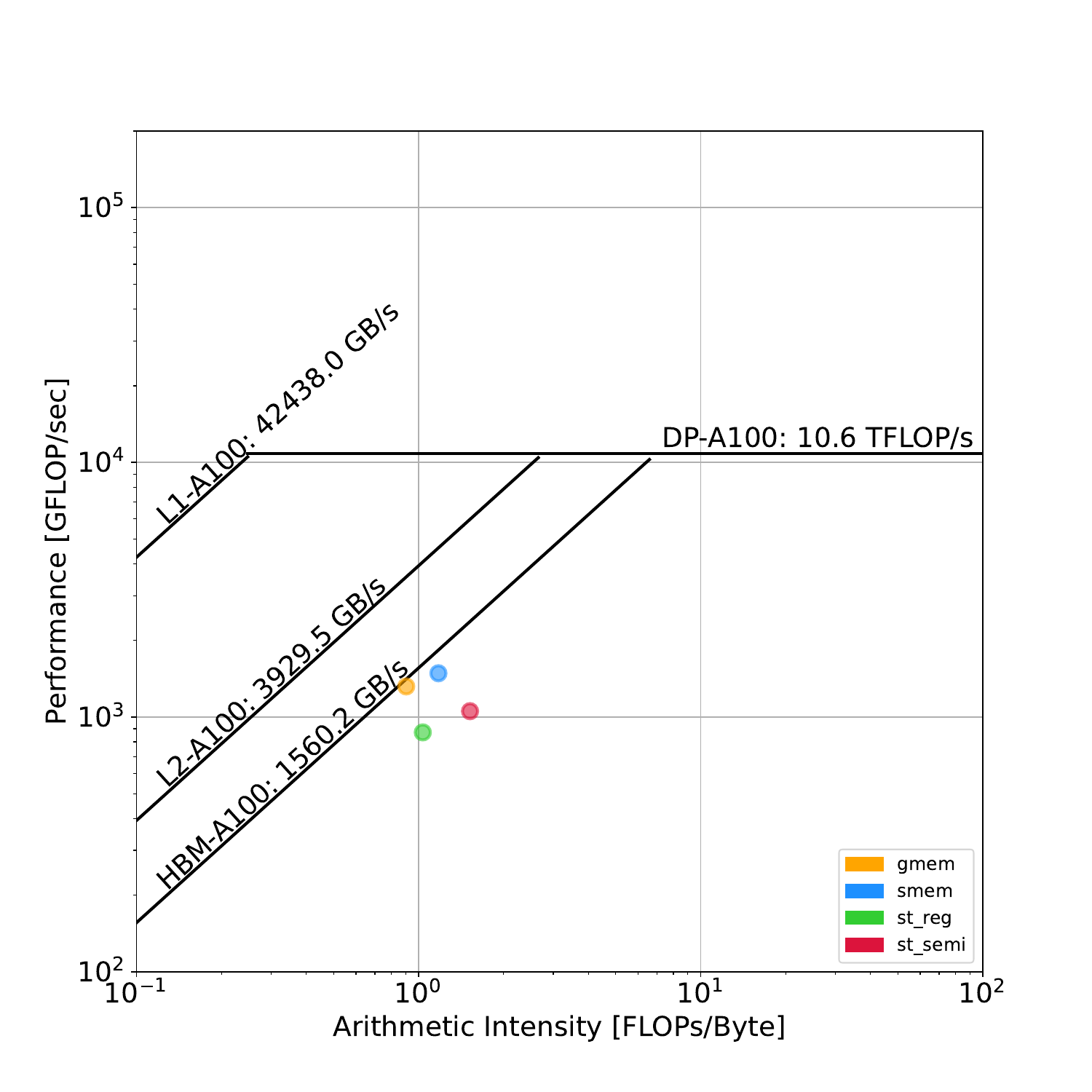}
  \caption{Roofline plot of CUDA kernels on A100}
  \label{fig:a100-cuda-roofline}
\end{figure}

\begin{figure}
\centering
\includegraphics[width=\linewidth]{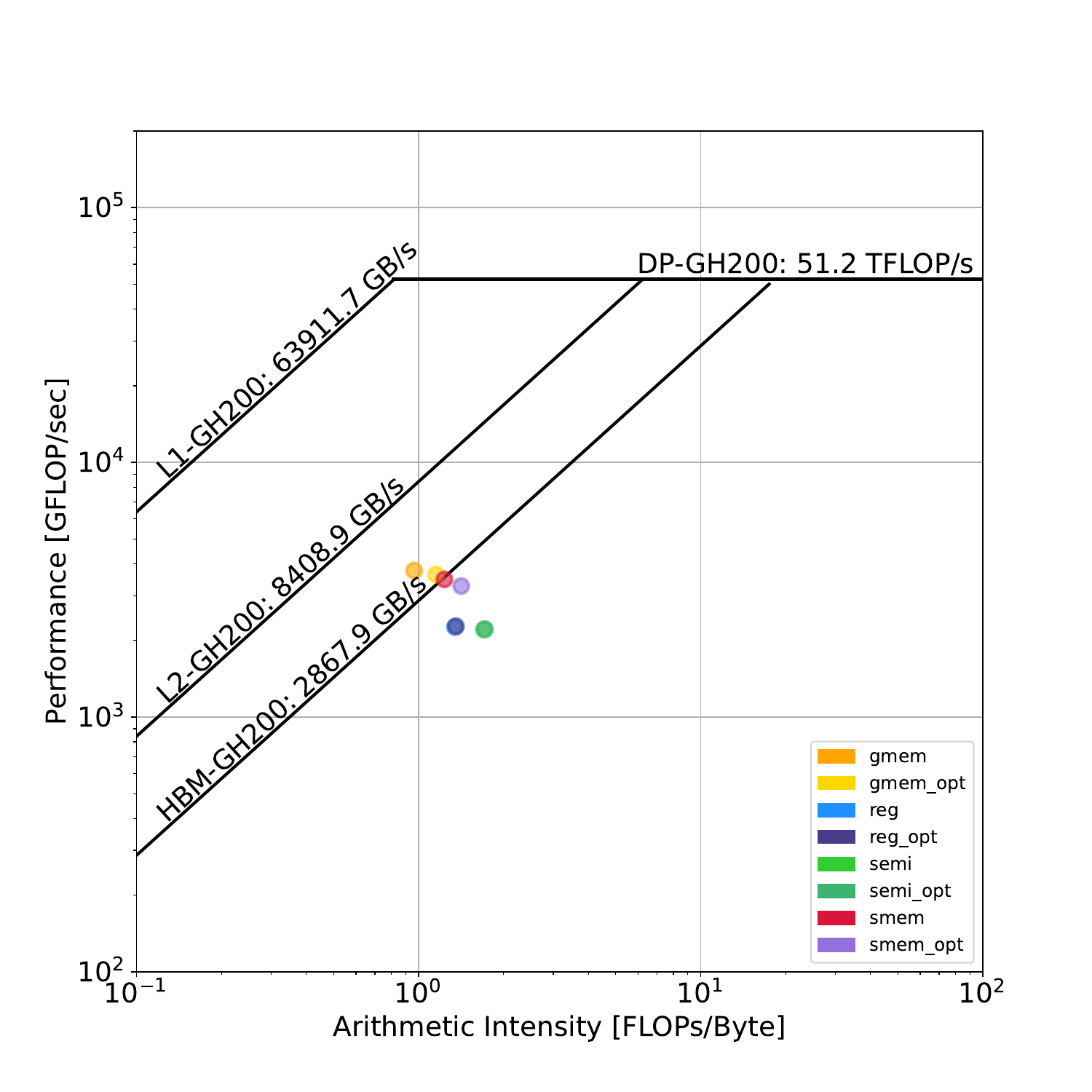}
  \caption{Roofline plot of CUDA kernels on GH200}
  \label{fig:h100-cuda-roofline}
\end{figure}


Finally, \autoref{fig:a100-cuda-roofline} and ~\autoref{fig:h100-cuda-roofline} show the roofline plot of the CUDA kernels on A100 and GH200. In order to avoid confusion due to large amount of information in the roofline model graph, the information per kernel represent the average of the three different memory levels. In the roofline plot of GH200, the position of points is noticeably closer to the upper-left corner compared to the plot of A100. This signifies higher computational performance and efficiency. With the analysis from profiling, it can be concluded that the memory performance enhancement brought by the Hopper architecture significantly aids memory-bound programs like stencil computation. Additionally, apart from the \textit{gmem} kernel, the other kernels and their corresponding optimized versions exhibit substantial overlap. This observation further substantiates the limited performance improvement brought about by thread block clustering for highly optimized code of this nature.

\subsection{Analysis and Recommendations on Performance Differences among Different Scenarios}

\subsubsection{CUDA kernel with and without thread block clusters on NVIDIA Hopper Architecture}
\label{sec:cluster}

As introduced in the NVIDIA Hopper white paper~\cite{h100}, the primary purpose of the new NVIDIA Hopper Architecture's thread block cluster is to optimize data transfers between adjacent thread blocks, thereby enhancing performance. This new feature positively impact should be mainly reflected in memory operations, not in computational performance. The results in ~\autoref{tab:cuda-prof-mem} and ~\autoref{tab:cuda-prof-comp} exactly confirm this point. According to~\autoref{tab:cuda-prof-mem}, it can be seen that for \textit{gmem}, \textit{reg}, and \textit{semi} kernels, when the thread block cluster is adopted, the read and write volume of the GPU memory significantly decreases, while the L1 and L2 cache hit rate increase.

The exception is the kernel using shared memory, where the L1 cache hit rate doesn't increase but decrease. But in this case, the benefits introduced by the improvement in GPU memory read and write operations are sufficient to lead this specific kernel to an enhanced performance.

However, overall, the introduction of thread block clusters in our CUDA kernel implementation for the stencil computation did not bring significant performance benefits. In our CUDA code, the shared memory located in the thread blocks is allocated and planned with fine granularity. We have minimized the data exchange between shared memory in different thread blocks, and almost every thread block performs computations only from its corresponding shared memory. Therefore, the benefits of thread block clusters are negligible.

\subsubsection{Optimal (``opt'') implementation and its contributing factors}
According to \autoref{cuda-time}, on NVIDIA Hopper Architecture, the optimal implementation is \textit{st\_reg\_fixed}, while for Ampere, the optimal one is \textit{st\_semi}.

The key factor behind this is the improved read and write management.
In \textit{st\_semi}, the advantage of the strategy lies in optimizing the balance between loading and storing. The data primarily resides in shared memory, and the benefits provided by the new generation GPU are mainly related to improvements in shared memory. On the other hand, \textit{st\_reg\_fixed} relies on data exchanges between registers, shared memory, and global memory. Therefore, based on the improvement observed in NVIDIA Hopper compared to NVIDIA Ampere as mentioned in ~\cite{h100}, there are significant improvements in shared memory size, memory read/write speed, and register size. As a result, \textit{st\_reg\_fixed} benefits more than \textit{st\_semi}.

From~\autoref{tab:cuda-prof-mem}, it can be observed that on the GH200, the memory read and write operations of \textit{st\_reg\_fixed} exhibit a decrease of 31.54\% compared to A100. However, this decrease is only 14.28\% in the case of \textit{st\_semi}, despite a similar improvement in throughput for both cases. This result explains why the performance of the \textit{st\_reg\_fixed} kernel is more advantageous on the GH200 as compared to the \textit{st\_semi} version. On GH200, the difference in performance due to differences about memory is even more pronounced.

Based on the analysis above, our recommendation for CUDA developers is to follow 
a strategy that fully exploits memory hierarchy, rather than only relying on block clustering, unless their algorithm has obvious locality patterns to be exploited.

\section{Exploration and Optimization of OpenACC and OpenMP Target Offloading}
\label{sec:acc-omp}

In this section, we will present and analyze the evaluation results of our stencil computation program on two general parallel computing programming models, OpenACC and OpenMP. We will first demonstrate the different implementations of these two programming models and the new optimization strategies we have proposed for this stencil computation program. Then, we will provide the evaluation and analysis of the profiling on NVIDIA Ampere and NVIDIA Hopper Architectures.

\subsection{Implementations and Optimization on OpenACC and OpenMP Target Offloading}

We adopted the same data domain decomposition as shown in ~\Cref{fig:domain}, which is identical to that under the CUDA model. The unoptimized inner region implementations of the program in OpenACC and OpenMP are similar, both implementing parallelism at the outermost loop of the 3D data as shown in \Cref{fig:acc} and \Cref{fig:omp}. The OpenACC implementation uses the directive \texttt{\#pragma acc loop gang vector collapse(3)}, while the OpenMP target offloading implementation uses \texttt{\#pragma omp target teams distribute parallel for simd collapse(3)}. 

\begin{figure}
\begin{minted}[frame=lines]{c}
#pragma acc parallel
#pragma acc loop gang vector collapse(3)
for (llint i = x3; i < x4; ++i) {
    for (llint j = y3; j < y4; ++j) {
        for (llint k = z3; k < z4; ++k) {
            # Computation
        }
    }
}
\end{minted}
\caption{OpenACC code for inner region}
\label{fig:acc}
\end{figure}

\begin{figure}
\begin{minted}[frame=lines]{c}
#pragma omp target teams distribute \ 
    parallel for simd collapse(3) 
for (llint i = x3; i < x4; ++i) {
    for (llint j = y3; j < y4; ++j) {
        for (llint k = z3; k < z4; ++k) {
            # Computation
        }
    }
}
\end{minted}
\caption{OpenMP target offloading code for inner region}
\label{fig:omp}
\end{figure}

However, both of these implementations face issues of memory bandwidth bottlenecks and an inability to hide latency, problems which asynchronous computation can effectively address. In CUDA, we typically use the concept of streams to implement asynchronous instructions provided by the host. In OpenACC and OpenMP, we cannot manage streams with the same fine granularity as in CUDA. Alternatively, both OpenACC and OpenMP provide methods for implementing stream asynchronicity. For OpenACC, the \texttt{async} clause can be used to specify the CUDA stream ID. In the case of OpenMP, the \texttt{nowait} clause can be used to dispatch target tasks to the GPU, allowing the GPU to manage its own scheduling. In this process, each target task will be executed asynchronously.

\subsubsection{Runtime Optimization 1: Fine-Grained Concurrent Kernels}

In OpenACC, we adjust the parallel region from the outermost loop to the second layer loop, and then use the third layer loop to iterate over the stream IDs, thereby distributing the computation of the parallel region to different streams. This approach allows for different parallel regions to be executed asynchronously, particularly enabling memory read/write tasks and compute tasks from different parallel regions to effectively hide latency caused by insufficient bandwidth. However, adjusting the original parallel region from the third layer loop to the second layer loop means that the number of times the parallel region is executed, i.e., the number of CUDA kernels, has increased by several orders of magnitude compared to the original version (the specific increase depends on the size of the data). This increase in number implies an increase in kernel launch or creation of CUDA streaming overhead, thus posing a potential risk of performance degradation.

\begin{figure*}
\centering
\includegraphics[width=.95\textwidth]{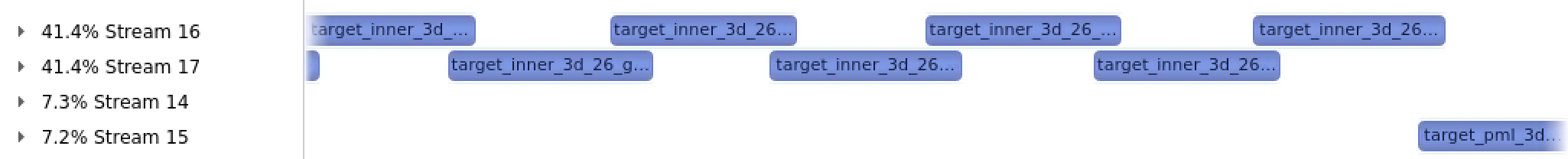}
  \caption{Tracing results for OpenACC (Grid Size: $1024^3$, 1000 iterations)}
  \label{fig:acc_prof}
\end{figure*}

\begin{figure*}
\centering
\includegraphics[width=.9\textwidth]{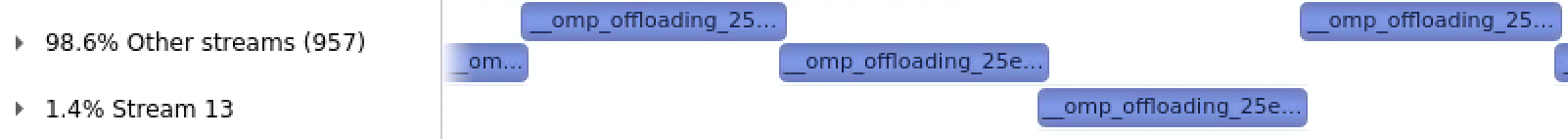}
  \caption{Tracing results for OpenMP target offloading (Grid Size: $1024^3$, 1000 iterations)}
  \label{fig:omp_prof}
\end{figure*}

\begin{figure}[h]
  \centering
  
\begin{tikzpicture}

\definecolor{darkgray176}{RGB}{176,176,176}
\definecolor{darkorange25512714}{RGB}{175,141,195}
\definecolor{lightgray204}{RGB}{204,204,204}
\definecolor{steelblue31119180}{RGB}{127,191,123}

\begin{axis}[
legend cell align={left},
legend style={
  fill opacity=0.8,
  draw opacity=1,
  text opacity=1,
  at={(0.97,0.03)},
  anchor=south east,
  draw=lightgray204
},
tick align=outside,
tick pos=left,
x grid style={darkgray176},
xlabel={Grid Size(\(\displaystyle x^3\))},
xmajorgrids,
xmin=55, xmax=1045,
xtick style={color=black},
y grid style={darkgray176},
ylabel={Grid/s},
ymajorgrids,
ymin=3025999.27347555, ymax=77500451.5905755,
ytick style={color=black}
]
\addplot [semithick, darkorange25512714, mark=square*, mark size=2, mark options={solid}]
table {%
100 6411201.65152554
125 8967968.99751594
150 11680908.7227835
175 15059415.7614041
200 18457463.6214928
225 21879339.8525963
250 25764271.3451835
275 30316718.8299487
300 34834439.6493333
325 42352166.4569299
350 44025247.6452268
375 49234308.0413411
400 54672350.3130846
425 59872110.3450427
450 63600577.9015473
475 66177946.216308
500 69042845.2280347
525 70375420.6871061
550 69656396.6657037
575 70625896.247093
600 71119071.7644115
625 70431410.7594133
650 71691675.5424681
675 72635027.4199719
700 72427045.0610245
725 73462967.6038825
750 73593417.1942133
775 73812479.0884051
800 74115249.2125255
825 73562283.839019
850 73348238.8659374
875 73043394.4716057
900 72064056.9395018
925 70720392.1796395
950 69134782.0828126
975 66828611.2392928
1000 64955732.6681866
};
\addlegendentry{OpenACC with \textit{async}}
\addplot [semithick, steelblue31119180, mark=x, mark size=3, mark options={solid}]
table {%
100 16410417.9897566
125 23813341.2015153
150 33700460.3233247
175 40804109.8184919
200 46707964.2917613
225 51511221.9564146
250 55292706.3169926
275 58360647.2270318
300 63366596.8851797
325 65526569.9653165
350 68731745.0088329
375 70570505.7690918
400 68714870.8643579
425 61077793.6905756
450 57529498.6647474
475 57785498.6115979
500 57840379.0627082
525 57671786.0729829
550 58223156.9811796
575 57447526.191577
600 56115264.9108131
625 56449221.4956901
650 56131146.0666808
675 56366185.9928925
700 56439165.8192564
725 56485390.9217979
750 56608900.4287181
775 56807538.0214448
800 56036879.2711703
825 56390149.2213556
850 56131635.7121966
875 55995275.3700716
900 56217900.2729923
925 56253109.5632396
950 56411445.7910597
975 56107326.8399579
1000 54425619.2274828
};
\addlegendentry{OpenACC without \textit{async}}
\end{axis}

\end{tikzpicture}
  \caption{Performance of OpenACC with and without \textit{async} on A100}
  \label{fig:acc-async}
\end{figure}
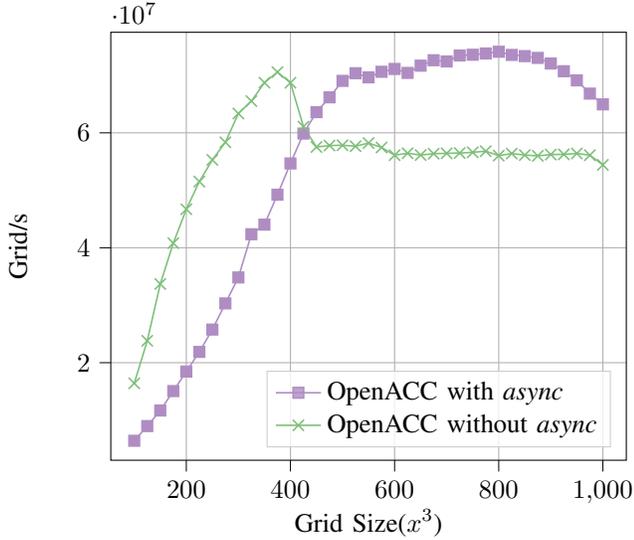

Due to the limited number of CUDA kernels that NVIDIA GPUs can execute simultaneously, excessive CUDA streaming does not lead to greater overlap, meaning it does not bring effective concurrency. Therefore, we limited the number of CUDA streams available in OpenACC to \textbf{2} and ensured that the CUDA Stream IDs between two adjacent kernels are different. In this way, we can reduce the overhead of launching streams while ensuring concurrency. The tracing results provided by the NVIDIA Nsight System is shown in~\autoref{fig:acc_prof}. It can be seen that the program mainly uses two CUDA streams to perform the inner part of the computation, while other streams handle boundary computation.

~\autoref{fig:acc-async} shows performance of the \texttt{async} version of OpenACC and the original version of OpenACC under different grid sizes. In order to better reflect the computing power of the stencil computation program under different grid sizes, we did not use the number of floating-point operations per second (FLOPS) as the evaluation standard, but chose the number of grids processed per second (grid/s) as the performance reference indicator. 
It can be seen that when the grid size is large, the original version reach the performance bottleneck and the throughput does not increase with the grid size. But the asynchronous execution of parallel regions can effectively hide the latency caused by memory read/write and kernel launch, thereby improving performance. 

In OpenMP target offloading, there is no directive equivalent to that in OpenACC that can specify CUDA streams. The asynchronous execution can only be indirectly achieved through the \texttt{nowait} clause. 

\begin{table*}[]
\centering
\caption{The kernel performance of asynchronous execution on the second layer loop in OpenMP and OpenACC.}
\label{tab:omp-acc-sync}
\begin{tabular}{@{}ccccc@{}}
\toprule
\multicolumn{1}{l}{} & \textbf{Kernel Time(us)} & \textbf{Cycles} & \textbf{Performance(GFLOPs)} & \textbf{Arithmetic Intensity} \\ \midrule
\textbf{OpenMP}               & 101.18          & 109660 & 381.254              & 0.88                 \\
\textbf{OpenACC}              & 35.74           & 38805  & 1079.255               & 0.88                 \\ \bottomrule
\end{tabular}
\end{table*}

Initially, the same code optimization as in the OpenACC version was used, i.e. reducing the target task to the second layer loop and making it execute asynchronously. However, compared with OpenACC, the reduced OpenMP kernel does not yield any performance improvement. On the contrary, we found that the same ``small kernel'' can be executed in 35us with OpenACC, while OpenMP requires more than 101us to complete. In terms of total kernel time, the OpenMP code using this optimization strategy has no improvement at all, but rather a significant decrease. 
The ~\autoref{tab:omp-acc-sync} shows the performance comparison between OpenMP kernel and OpenACC kernel under the same kernel. It can be seen that OpenMP target offloading shows significantly weaker performance when facing this relatively small kernel. 

On the one hand, this behavior mainly comes from the differences in how OpenACC and OpenMP handle the working mechanisms of CUDA kernels. In OpenACC, the kernel code region is compiled into machine code by the compiler and then directly run on the NVIDIA GPU. However, the OpenMP target offloading region includes some necessary device runtime when running on the NVIDIA GPU. In OpenMP, this runtime code includes but is not limited to kernel launching, fine-grained memory management, and potential multi-device support. Particularly, with regards to fine-grained memory management, OpenMP target offloading supports loading necessary data into the shared memory of the NVIDIA GPU, which OpenACC does not support. This runtime support provided by OpenMP can effectively improve the program's running efficiency in some cases, but it often backfires for the small kernels in this optimization scheme. This is because the additional runtime overhead on the GPU is hard to be hidden by the kernel code, and thus the runtime overhead becomes an unavoidable part of the execution time.

On the other hand, OpenMP target offloading has a non-negligible overhead for the creation and deletion of CUDA streams. Since the \texttt{nowait} clause cannot specify the CUDA stream ID, OpenMP target offloading can create at most as many CUDA streams as there are kernels. \autoref{fig:omp_prof} shows the tracing result of the OpenMP target offloading program in the Nsight System. It can be seen that the program has launched nearly 1000 CUDA streams, which brought significant overhead.


\subsubsection{Runtime Optimization 2: Coarse-Grained Concurrent Kernels}

The second performance optimization approach is to keep the original size of the target task unchanged, and make the inner loop and the boundary task execute asynchronously. Since the boundary task is significantly smaller than the inner task, the performance improvement yield by this optimization method is relatively small. 

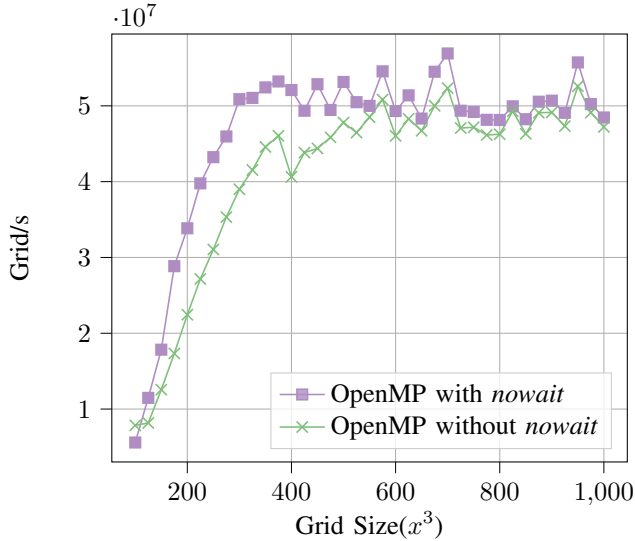
\begin{figure}[h]
\centering
\begin{tikzpicture}

\definecolor{darkgray176}{RGB}{176,176,176}
\definecolor{steelblue31119180}{RGB}{175,141,195}
\definecolor{lightgray204}{RGB}{204,204,204}
\definecolor{darkorange25512714}{RGB}{127,191,123}

\begin{axis}[
legend cell align={left},
legend style={
  fill opacity=0.8,
  draw opacity=1,
  text opacity=1,
  at={(0.97,0.03)},
  anchor=south east,
  draw=lightgray204
},
tick align=outside,
tick pos=left,
x grid style={darkgray176},
xlabel={Grid Size(\(\displaystyle x^3\))},
xmajorgrids,
xmin=55, xmax=1045,
xtick style={color=black},
y grid style={darkgray176},
ylabel={Grid/s},
ymajorgrids,
ymin=3021254.93210605, ymax=59476529.0791192,
ytick style={color=black}
]
\addplot [semithick, steelblue31119180, mark=square*, mark size=2, mark options={solid}]
table {%
100 5587403.75697029
125 11480124.1396336
150 17832517.0003329
175 28849362.925322
200 33847963.4103516
225 39773957.3439857
250 43230217.1609437
275 45960655.7449508
300 50880515.5892246
325 51049106.7777228
350 52434232.4912467
375 53215764.8084573
400 52083757.1920344
425 49355535.0529781
450 52854889.0409847
475 49462953.0486313
500 53153264.248264
525 50484467.7265734
550 49996994.9214172
575 54552944.7380956
600 49313267.0952659
625 51384287.7798989
650 48319782.3872304
675 54497990.526769
700 56910380.254255
725 49357592.3872779
750 49194808.5265171
775 48158072.9295989
800 48134777.3766546
825 49936913.6022055
850 48235899.3692908
875 50550603.6596869
900 50683074.356033
925 49083581.9183111
950 55729068.4901233
975 50231107.636612
1000 48467224.0397431
};
\addlegendentry{OpenMP with \textit{nowait}}
\addplot [semithick, darkorange25512714, mark=x, mark size=3, mark options={solid}]
table {%
100 7848677.49784161
125 8164419.12182724
150 12560803.5936924
175 17334158.0956077
200 22445366.5749212
225 27160889.0351379
250 31051703.8194093
275 35330380.1984235
300 38996490.3158716
325 41542675.2378578
350 44582371.0646634
375 46050592.0673455
400 40634662.6370626
425 43830255.8480784
450 44384750.6417219
475 45848734.3369654
500 47797126.0544046
525 46466629.5241368
550 48509225.1352864
575 50827170.742532
600 46037964.2694244
625 48281386.5783864
650 46710889.9944721
675 50007865.8920287
700 52334452.2429051
725 47069577.6711141
750 47173819.9416526
775 46150161.6053459
800 46256979.2024285
825 49297270.0695322
850 46319342.3087076
875 49092552.0844783
900 49131935.0838411
925 47329202.6216489
950 52587111.0593171
975 49101750.6079051
1000 47219480.8690273
};
\addlegendentry{OpenMP without \textit{nowait}}
\end{axis}

\end{tikzpicture}
  \caption{Performance of OpenMP Target Offloading with and without \textit{nowait} on A100}
  \label{fig:omp-nowait}
\end{figure}

The ~\autoref{fig:omp-nowait} shows the changes in execution time of the original version and the asynchronous version under different grid sizes. Unlike the optimization approach of reducing the kernel to make it execute asynchronously in OpenACC, this optimization method does not change the kernel itself. Therefore, in subsequent discussions, we will only discuss the optimized kernel and no longer distinguish between the two versions of the kernel.

\subsubsection{Compilation Optimization}

\begin{table*}[]
\centering
\caption{Register spills and performance differences caused by different register limits in OpenACC}
\label{tab:acc-reg}
\resizebox{0.6\textwidth}{!}{%
\begin{tabular}{@{}cccccc@{}}
\toprule
\textbf{Max Register} & \textbf{Stack Frame} & \textbf{Spill Stores} & \textbf{Spill Loads} & \textbf{$1024^3$} & \textbf{$512^3$} \\ \midrule
No Setup                & 0                    & 0                     & 0                    & 9.97122 s         & 1.12388 s        \\
100                     & 0                    & 0                     & 0                    & 9.96777 s         & 1.12309 s        \\
80                      & 0                    & 0                     & 0                    & 9.60158 s         & 1.03545 s        \\
60                      & 8                    & 8                     & 8                    & \textbf{9.55082 s}         & \textbf{0.954919 s}       \\
40                      & 72                   & 68                    & 68                   & 13.7608 s         & 1.61103 s        \\ \bottomrule
\end{tabular}}

\end{table*}

In addition, we further improved the performance of OpenACC programs by optimizing the number of registers. The number of registers affects the performance of GPU programs. On the one hand, excessive use of registers may limit the number of parallel threads, thereby affecting performance; on the other hand, when there are not enough registers to store all variables, some variables need to be stored in local memory (which is part of global memory), which we refer to as spillover storage and loading. Excessive spillover storage and loading will also lead to a decrease in performance. \autoref{tab:acc-reg} shows the impact of different register caps on the performance of OpenACC-\textit{async} programs.

\subsection{Performance and Profiling Results on OpenACC and OpenMP}

\begin{table}[h]
    \centering
    \caption{Execution time of OpenACC and OpenMP programs on A100 and GH200 (Grid Size: $1024^3$, 1000 iterations)}
    \label{tab:acc-omp-time}
\resizebox{.8\linewidth}{!}{%
\begin{tabular}{@{}ccc@{}}
\toprule
\textbf{Programming Model }             & \textbf{Device} & \textbf{Execution Time(s)} \\ \midrule
\multirow{2}{*}{OpenACC}       & A100   & 53.188           \\
                               & GH200   & 23.196           \\ \midrule
\multirow{2}{*}{OpenACC-async} & A100   & 44.222            \\
                               & GH200   & 19.229           \\ \midrule
\multirow{2}{*}{OpenMP}        & A100   & 58.568           \\
                               & GH200   & 29.527           \\ \bottomrule
\end{tabular}}

\end{table}

\autoref{tab:acc-omp-time} shows execution time for OpenACC, OpenACC-\textit{async}, OpenMP on A100 and GH200. The grid size used in the experiments is $1024^3$, and the kernel allocates in total about 22.1 GB GPU memory. The number of the experiments time steps is 1000. 

\begin{table*}[!h]
\centering
\caption{Profiling results of memory information of OpenACC and OpenMP code on NVIDIA A100 and GH200 GPUs.}
\label{tab:acc-omp-prof-memory}
\resizebox{\textwidth}{!}{%
\begin{tabular}{@{}cccrccr@{}}
\toprule
\textbf{Kernel}                & \textbf{Device} & \textbf{Memory Throughput Rate} & \multicolumn{1}{c}{\textbf{Memory Throughput(GB/s)}} & \textbf{L1 Hit Rate} & \textbf{L2 Hit Rate} & \multicolumn{1}{c}{\textbf{Device Memory Read/Write(Bytes)}} \\ \midrule
\multirow{2}{*}{OpenACC}       & A100            & 60.95\%                         & 947                                                  & 42.97\%              & 55.16\%              & 47,612,893,312                                               \\
                               & GH200           & 60.22\%                         & 2,422                                                & 43.44\%              & 45.31\%              & 45,280,361,216                                               \\ \midrule
\multirow{2}{*}{OpenACC-async} & A100            & 56.18\%                         & 872                                                  & 48.93\%              & 48.41\%              & 46,209,280                                                   \\
                               & GH200           & 52.26\%                         & 2,097                                                & 46.13\%              & 43.74\%              & 44,710,400                                                   \\ \midrule
\multirow{2}{*}{OpenMP}        & A100            & 54.39\%                         & 845                                                  & 47.04\%              & 54.51\%              & 46,086,569,472                                               \\
                               & GH200           & 30.03\%                         & 1,208                                                & 46.01\%              & 56.66\%              & 32,045,994,240                                               \\ \bottomrule
\end{tabular}
}
\end{table*}
\begin{table*}[h]
\centering
\caption{Profiling results of compute information of OpenACC and OpenMP code on NVIDIA A100 and GH200 GPUs.}
\label{tab:acc-omp-prof-comp}
\resizebox{\textwidth}{!}{%
\begin{tabular}{@{}cccrrcccc@{}}
\toprule
\multirow{2}{*}{\textbf{Kernel}} & \multirow{2}{*}{\textbf{Device}} & \multicolumn{1}{c}{\multirow{2}{*}{\textbf{Compute Throughput}}} & \multirow{2}{*}{\textbf{SM Frequency}} & \multirow{2}{*}{\textbf{Elapsed Cycles}} & \multicolumn{2}{c}{\textbf{Theoretical Occupancy}}                      & \multicolumn{2}{c}{\textbf{Achieved}}                                   \\ \cmidrule(l){6-9} 
                                 &                                  & \multicolumn{1}{c}{}                                             &                                        &                                          & \multicolumn{1}{c}{Occupancy} & \multicolumn{1}{c}{Active Warps Per SM} & \multicolumn{1}{c}{Occupancy} & \multicolumn{1}{c}{Active Warps Per SM} \\ \midrule
\multirow{2}{*}{OpenACC}         & A100                             & 40.87\%                                                          & 764,999,157.62                         & 38,427,322                               & 37                            & 24                                      & 37                            & 24                                      \\
                                 & GH200                            & 45.15\%                                                          & 1,529,167,787.04                       & 28,593,676                               & 37                            & 24                                      & 37                            & 24                                      \\ \midrule
\multirow{2}{*}{OpenACC-async}   & A100                             & 47.21\%                                                          & 765,073,370.74                         & 40,523                                   & 31                            & 20                                      & 27                            & 17                                      \\
                                 & GH200                            & 48.40\%                                                          & 1,508,094,031.53                       & 32,408                                   & 37                            & 24                                      & 32                            & 20                                      \\ \midrule
\multirow{2}{*}{OpenMP}          & A100                             & 42.83\%                                                          & 764,998,995.17                         & 41,677,024                               & 25                            & 16                                      & 24                            & 15                                      \\
                                 & GH200                            & 58.30\%                                                          & 1,529,855,754.88                       & 40,585,290                               & 25                            & 16                                      & 23                            & 14                                      \\ \bottomrule
\end{tabular}}
\end{table*}

\autoref{tab:acc-omp-prof-memory} and ~\autoref{tab:acc-omp-prof-comp} show the profiling of the memory information and computation information respectively. Same as ~\autoref{cuda-time}, these two profiling results are based on grid size of $1024^3$ and 1000 iterations. 

\subsection{Analysis of OpenACC and OpenMP on A100 and GH200}
The performance gap between OpenMP and OpenACC originates both from memory management and from differences in computational efficiency. As shown in~\autoref{tab:acc-omp-prof-memory}, the memory throughput of OpenMP is significantly less than that of the OpenACC version, and yet the volume of memory read/write operations is not less than that of the OpenACC version, thus, memory read/write operations will introduce significant latency. As shown in~\autoref{tab:acc-omp-prof-comp}, the elapsed cycles of OpenMP kernel are also significantly higher than the OpenACC version.

In addition, in OpenACC, the program can relatively automatically configure GPU-related settings, such as automatically adjusting the GPU's grid size according to the size of the data. OpenMP developers can manually select the grid size through \texttt{num\_teams}. However, compared to OpenACC's automatic setting of Grid Size, choosing a Grid Size is difficult, and the optimal value often changes with the size of the data. In~\autoref{fig:acc-async}, the curve of OpenACC is smoother than the curve of OpenMP in~\autoref{fig:omp-nowait}, which could prove that OpenACC can make fuller and more reasonable use of GPU resources.

\section{Programming Models Comparison}
\label{sec:gene}
\begin{table}[]
\centering
\caption{Lines of code of different GPGPU programming models implemented kernels}
\label{tab:loc}
\resizebox{.7\linewidth}{!}{%
\begin{tabular}{cc}
\hline
\multicolumn{1}{l}{\textbf{Programming Model}} & \multicolumn{1}{l}{\textbf{Lines of Code}} \\ \hline
CUDA-gmem                                      & 138                                         \\
CUDA-smem                                      & 145                                         \\
CUDA-st\_reg\_fixed                            & 142                                        \\
CUDA-st\_semi                                  & 172                                        \\
OpenACC                                        & 71                                         \\
OpenMP                                         & 69                                         \\ \hline
\end{tabular}}
\end{table}

\subsection{Portability of Different Programming Models}

In terms of performance alone, within the NVIDIA GPU-based programming models, the CUDA model, thanks to its fine-grained control over instructions and memory, as well as the vendor-specific implementation of certain instructions (such as \texttt{\_\_fmaf\_rn}), is far ahead when optimized. However, the development of scientific computing programs needs to take other costs into consideration, where portability and development difficulty cannot be ignored. ~\autoref{tab:loc} shows the line counts of different kernel implementations, and it can be seen that OpenACC and OpenMP Target Offloading far surpass the CUDA model in terms of portability. 
At the same time, considering that OpenMP Target Offloading code supports a variety of accelerators, we believe that the directive-based OpenMP Target Offloading programming model would be an excellent choice for developers of high-performance computing programs.

\subsection{Exploration of Performance of Different Generations of NVIDIA GPUs on Multiple Programming Models}

\begin{table}[]
\centering
\caption{Execution of OpenACC and OpenMP code on three generations of NVIDIA GPGPUs}
\label{tab:gene}
\resizebox{\linewidth}{!}{%
\begin{tabular}{@{}lrrr@{}}
\toprule
                          & \multicolumn{1}{l}{\textbf{Turing}} & \multicolumn{1}{l}{\textbf{Ampere}} & \multicolumn{1}{l}{\textbf{Hopper}} \\ \midrule
\textbf{CUDA }                     & 31.775s                & 5.706s                  & 3.364s                  \\
\textbf{OpenACC   }                & 81.718s                & 11.260s                  & 6.276s                  \\
\textbf{OpenMP  }                  & 139.358s                & 18.012s                  & 9.651s                 \\ \midrule
\textbf{(OpenACC - CUDA)/OpenACC } & 0.611              & 0.493                 & 0.464                  \\
\textbf{(OpenMP - CUDA)/OpenMP}    & 0.772               & 0.683                 & 0.651                 \\
\textbf{(OpenACC - OpenMP)/OpenMP} & 0.414               & 0.375                 & 0.349                  \\ \bottomrule
\end{tabular}}
\end{table}

In addition to the NVIDIA Ampere(A100) and Hopper Architecture(GH200), we evaluate the performance of three models on the NVIDIA Turing Architecture(T4) and compared the execution time for 1000 iterations when the grid size is $700^3$, the results of which can be seen in~\autoref{tab:gene}. We also calculated the performance differences of different models on different generations of GPUs and made some interesting findings. As can be seen from the last three lines of~\autoref{tab:gene}, as new generations of GPGPUs are introduced, the gap between the three programming models is gradually narrowing. Compared to the Turing, the performance of the OpenACC version on the Hopper has increased by $13.0\times$, while the performance of the OpenMP version has even achieved an $14.4\times$ increase; the optimized CUDA version, however, only saw a $9.4\times$ improvement with respect Turing reference.

Although the performance of OpenMP target offloading and OpenACC is $3\times$ and $2\times$ the optimized CUDA version, this is a result obtained at the expense of portability. The execution time of the optimized CUDA version in the chart is selected from a massive number of CUDA implementations, and selecting the optimal CUDA version requires a costly grid-like search, further in terms of code length or debugging difficulty. Therefore, we believe that OpenMP and OpenACC, whose performance continues to improve along with newer GPGPUs generations and compiler progress, are becoming development options worth considering when portability is a must.

In terms of architectural comparison, the rate of improvement (ri), defined as 
\begin{equation}
ri(Arch_{i+1}|Arch_{i}, prog\_model) = \frac{time\  Arch_{i}}{time\  Arch_{i+1}}
\end{equation}
was remarkable for ri(Ampere$|$Turing, CUDA) = 5.6, but less pronounced for ri(Hopper$|$Ampere, CUDA) = 1.7. The highest ri in Table~\ref{tab:gene} is achieved by OpenMP between Turing and Ampere architectures (7.7).
\section{Comparsion of Power Consumption}

\begin{figure*}[h]
    \centering
    \begin{subfigure}[b]{0.32\textwidth}
        \centering
        \includegraphics[width=\textwidth]{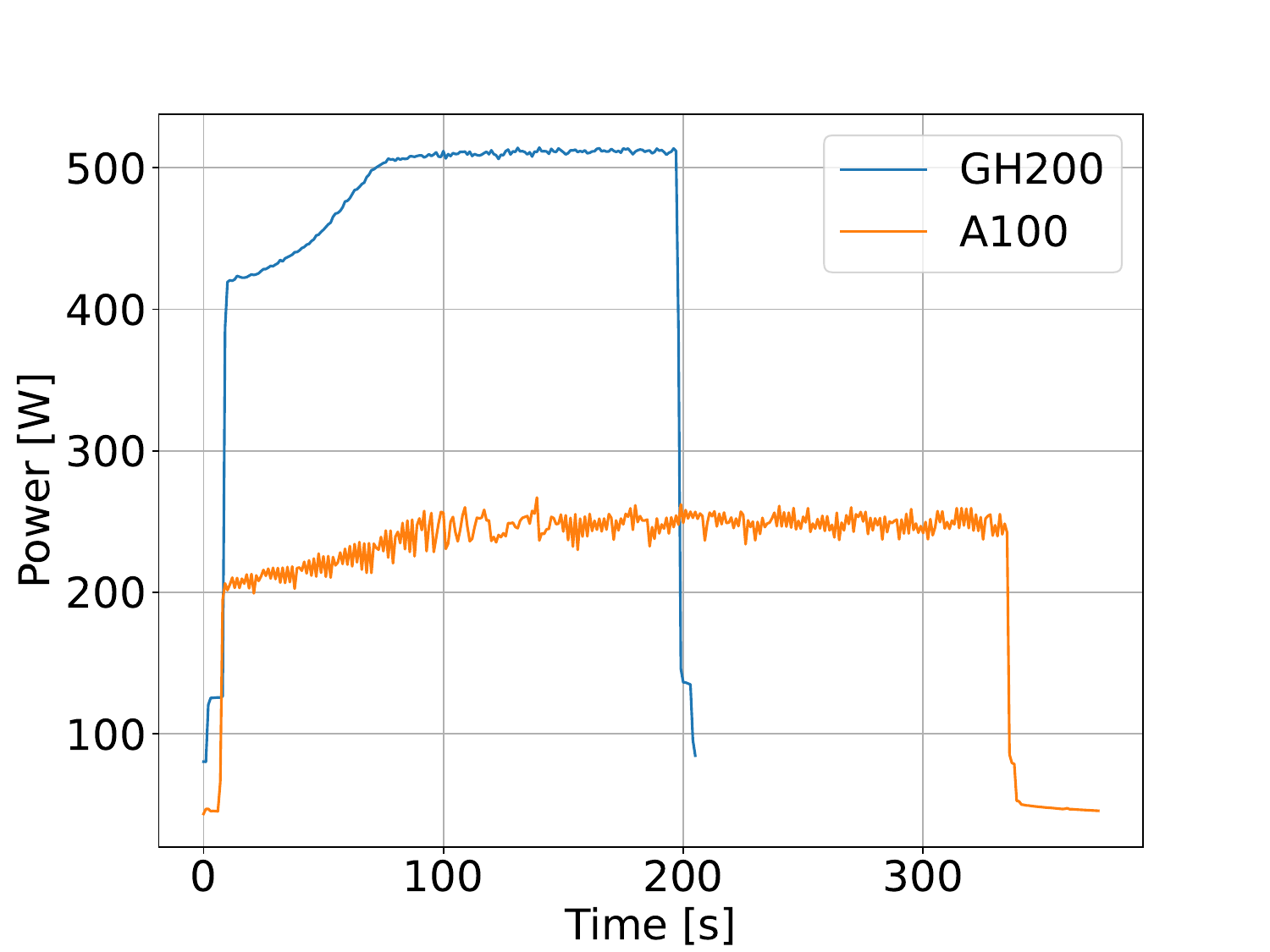}
        \caption{CUDA (optimal kernel)}
        \label{fig:sub1}
    \end{subfigure}
    \hfill
    \begin{subfigure}[b]{0.32\textwidth}
        \centering
        \includegraphics[width=\textwidth]{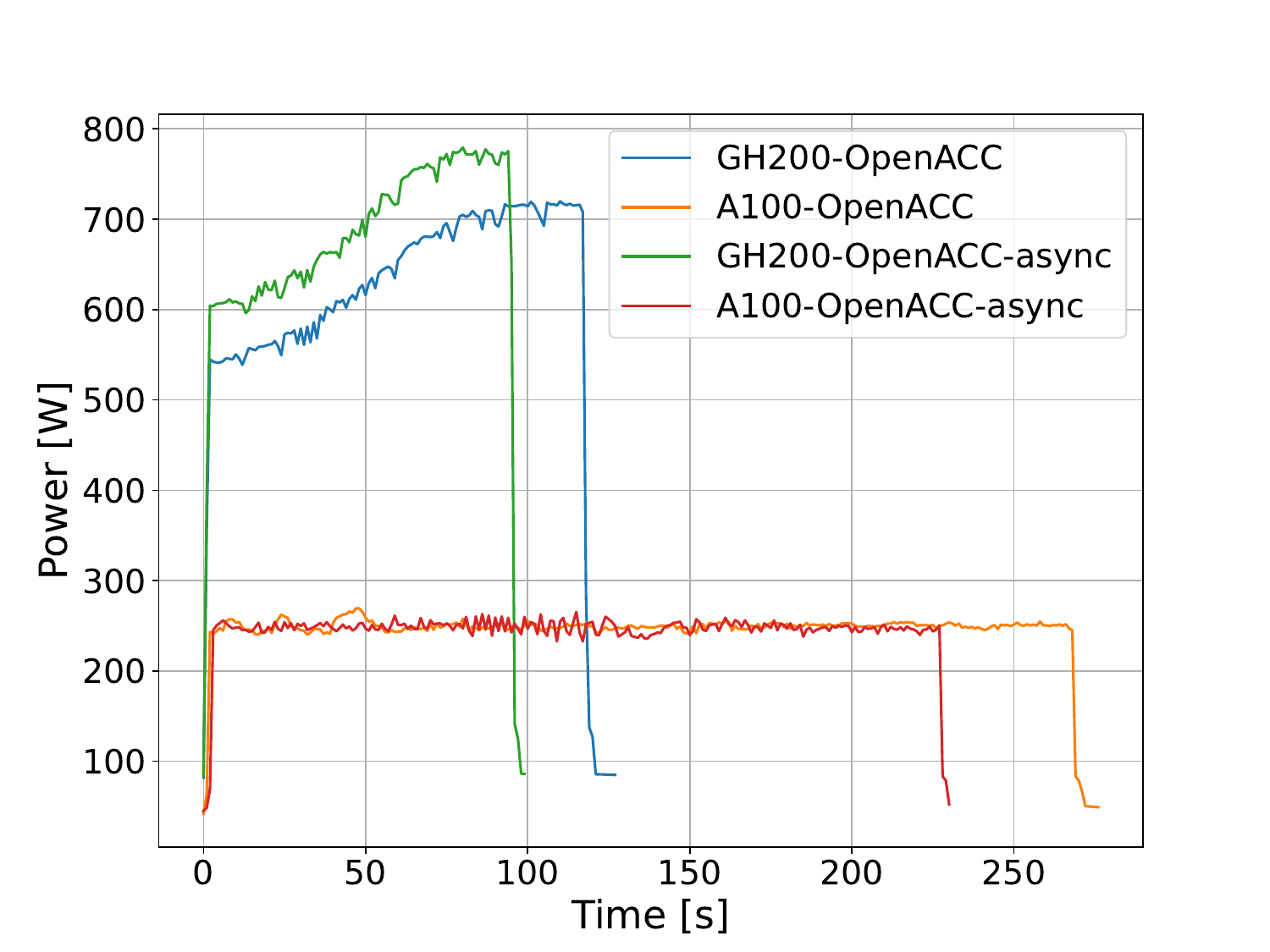}
        \caption{OpenACC}
        \label{fig:sub2}
    \end{subfigure}
    \hfill
    \begin{subfigure}[b]{0.32\textwidth}
        \centering
        \includegraphics[width=\textwidth]{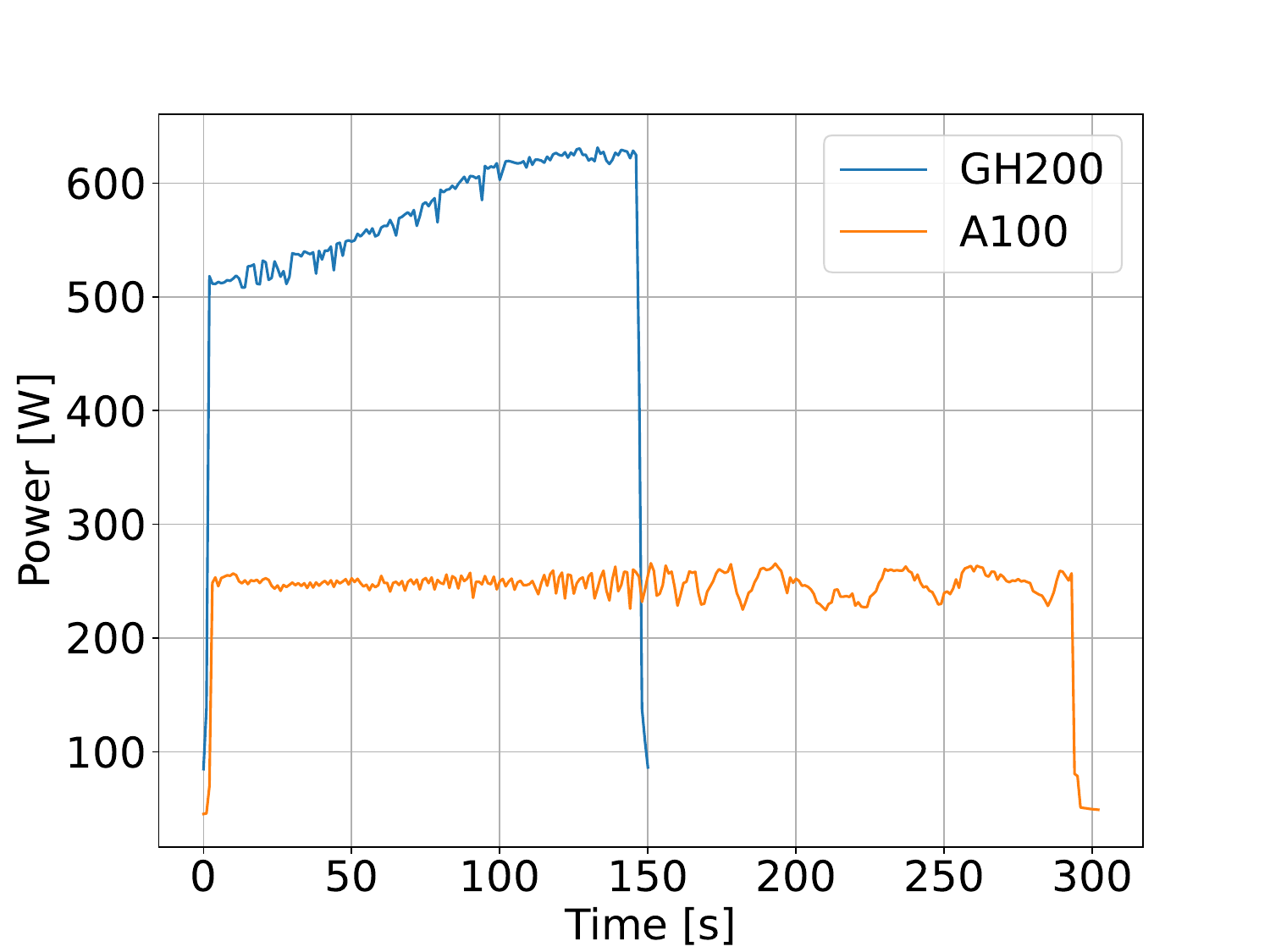}
        \caption{OpenMP}
        \label{fig:sub3}
    \end{subfigure}
    \caption{Power consumption of NVIDIA Ampere and NVIDIA Hopper across different programming models}
    \label{fig:power}
\end{figure*}

In this section, we evaluated the power consumption of NVIDIA Ampere Architecture and NVIDIA Hopper Architecture. \autoref{fig:power} shows the energy consumption curves of two different GPGUs across three distinct programming models. The grid size of this evaluation is $1024^3$, and the number of timesteps for CUDA is 10000, and for OpenACC/OpenMP is 5000.

The collected data provided insightful revelations regarding the energy efficiency of the A100 and GH200 GPUs in different programming models. Initial observations revealed that the GH200 GPU, while not exhibiting superior energy efficiency, tends to consume more energy for equivalent computational tasks compared to the A100. This increased energy consumption is particularly evident in OpenMP offloading implementations, where the GH200's architecture, 
it seems does not prioritize power efficiency. However, a notable advantage of the GH200 is its time to solution efficiency, where it significantly reduces computing time for similar tasks.

On the CUDA programming model side, both GPUs showed a significant increase in power consumption during the peak computational phases of the stencil computations. However, the GH200 GPU maintained a more stable energy profile.

During the OpenACC power consumption evaluation (Figure ~\ref{fig:power}.b), unlike its behavior with CUDA implementations, the GH200 exhibited greater fluctuations in energy consumption compared to the A100. This observation suggests that NVIDIA's newer Hopper architecture may not be as refined in energy management for OpenACC as it is for Ampere. Simultaneously, it was noted that the performance of the A100 remained consistently at its peak across both CUDA and OpenACC implementations. However, the GH200 only approached its theoretical performance peak with the OpenACC-async version. This indicates that CUDA's computational optimizations are highly advanced, and memory bandwidth has become the primary limiting factor in program performance.

The outcomes observed with OpenMP (Figure~\ref{fig:power}.c) are akin to those witnessed with CUDA (Figure~\ref{fig:power}.a), wherein the energy consumption fluctuations of the GH200 are relatively stable.
Overall, plenty of room for the application and SW stack to improve and take better advantage of the Hopper architecture. For instance, for the CUDA implementation, time to solution reduced from 340s to 200s ($1.7\times$) but the power consumption raised from 250W to 510W ($2\times$), so at least $0.3\times$ (power to time solution ratio) to catch-up.

\section{Conclusion}
Performance evaluation of optimized 3D stencil kernels was conducted on three GPU programming models for the latest generation GPGPUs, providing optimization suggestions tailored to each programming model. Our findings indicate that the GH200 demonstrates performance improvements of up to 58\% compared to the previous GPU generation. 

Simultaneously, we proposed CUDA stream-based asynchronous execution strategies for the OpenACC and OpenMP of stencil computation programs, resulting in performance enhancements of up to 30\% over the original versions.

We also compared the performance and portability of the three programming models on multiple GPGPU generations. Our observations reveal that as GPU architecture and performance progress, the performance gap between OpenMP, OpenACC, and optimized CUDA versions narrows. Particularly, in scenarios demanding portability, OpenMP and OpenACC are gradually becoming viable alternatives.

\bibliographystyle{IEEEtranS}
\bibliography{pmbs}

\end{document}